\documentclass[12pt]{article}
\pdfoutput=1

\usepackage{epsfig}
\usepackage{amsfonts}
\usepackage{amscd}
\usepackage{latexsym}
\usepackage{amsmath,amssymb}
\usepackage{verbatim}
\usepackage{setspace}
\usepackage{color}
\usepackage{fancyhdr}
\usepackage{cite}
\usepackage{hyperref}
\usepackage{tikz}
\usepackage{slashed}
\usetikzlibrary{calc}   
\usetikzlibrary{topaths}
\usetikzlibrary{decorations}
\usetikzlibrary{decorations.pathmorphing}

\usepackage{draft} 
\usepackage{hyperref}
\usepackage{graphicx,color,subfig}
\usepackage{cite}
\usepackage{mciteplus}
\usepackage{skak}
\usepackage[version=3]{mhchem}
\usepackage{empheq}

\DeclareFontFamily{OT1}{pzc}{}
\DeclareFontShape{OT1}{pzc}{m}{it}{<-> s * [1.10] pzcmi7t}{}
\DeclareMathAlphabet{\mathpzc}{OT1}{pzc}{m}{it}

\numberwithin{equation}{section}

 \def\p{\partial}
 \def\bz{{\bar z}}
 
\def\0{{(0)}}
\def\1{{(1)}}
\def\2{{(2)}}

\def\<{\langle }
\def\>{\rangle }
\def\bw{{\bar w}}

\def\eads${H$_3$}

\newcommand{\ba}{\begin{align}}
\newcommand{\ea}{\end{align}}
\def\be{\begin{equation}}
\def\ee{\end{equation}}
\def\beq{\be\begin{array}{c}}
\def\eeq{\end{array}\ee}

\def\be#1\ee{\begin{align}#1\end{align}}

\begin{document}

\newcommand{\scriplus}{\mathcal{I}^+}

\newcommand{\Del}{\nabla}
\newcommand\scalemath[2]{\scalebox{#1}{\mbox{\ensuremath{\displaystyle #2}}}}

	 \renewcommand{\theequation}{\thesection.\arabic{equation}}
   \makeatletter
  \let\over=\@@over \let\overwithdelims=\@@overwithdelims
  \let\atop=\@@atop \let\atopwithdelims=\@@atopwithdelims
  \let\above=\@@above \let\abovewithdelims=\@@abovewithdelims
\renewcommand\section{\@startsection {section}{1}{\z@}%
                                   {-3.5ex \@plus -1ex \@minus -.2ex}
                                   {2.3ex \@plus.2ex}%
                                   {\normalfont\large\bfseries}}

\renewcommand\subsection{\@startsection{subsection}{2}{\z@}%
                                     {-3.25ex\@plus -1ex \@minus -.2ex}%
                                     {1.5ex \@plus .2ex}%
                                     {\normalfont\bfseries}}

\renewcommand{\H}{\mathcal{H}}
\newcommand{\SU}{\mbox{SU}}
\newcommand{\chiu}{\chi^{{\rm U}(\infty)}}
\newcommand{\ff}{\rm f}
\linespread{1.3}

\unitlength = .8mm

\begin{titlepage}

\begin{center}

\hfill \\
\hfill \\
\vskip 1cm


\title{A Note on the Subleading Soft Graviton}

\author{Elizabeth Himwich,$^1$ Zahra Mirzaiyan,$^{2,3}$ and Sabrina Pasterski$^1$}

\address{
$^1$Center for the Fundamental Laws of Nature, Harvard University,\\
Cambridge, MA 02138, USA
\\
$^2$Physics Department, Isfahan University of Technology,\\
 84156-83111, Isfahan, Iran\\
$^3$Erwin Schr\"odinger Institute for Mathematics and Physics, \\ 
Boltzmanngasse 9A, 1090 Wien, Austria

}

\end{center}

\vspace{2.0cm}

\begin{abstract}

We show that the soft part of the charge generating infinitesimal superrotations can be expressed, in harmonic gauge, in terms of metric components evaluated at the boundaries of null infinity that are subleading in a large radius expansion. We then recast the spin memory observable in terms of these boundary values.

\end{abstract}

\vfill

\end{titlepage}

\eject

\tableofcontents

\section{Introduction} 
Recently,~\cite{Himwich:2019dug} identified a new celestial current corresponding to Low's subleading soft theorem in electromagnetism.   
 There, the current could be expressed in terms of the boundary values of the gauge potential at null infinity, but at one order subleading in a large radius expansion compared to the current~\cite{He2014} corresponding to the leading soft theorem.  In this note, we examine the analogous computations for gravity in harmonic gauge.  We find that the subleading soft graviton mode that appears in the superrotation charge~\cite{Kapec2014}, the 2D stress tensor for 4D gravity~\cite{Kapec2017}, and the spin memory observable~\cite{Pasterski2016} can also be neatly recast in terms of a difference between boundary values of the metric.  The relevant metric component is $h_{zz}^{(0)}$, which is subleading to the radiative data where the superrotation Goldstone mode appears, $h_{zz}^{(-1)}$.  We hope that rewriting the generator of inhomogeneous shifts in the superrotation Goldstone mode in terms of a boundary difference of $h_{zz}^{(0)}$ will help to bridge the gap between our understanding of the leading supertranslation example~\cite{He:2014laa} and recent studies of the $\Delta=2$ Goldstone mode in~\cite{Donnay:2018neh}.

\section{Setup}

We consider linearized gravity in four dimensions. In this section we set up Einstein's equations in harmonic gauge, impose boundary conditions on the metric perturbations, and identify the residual symmetries allowed by these boundary conditions. 

\subsection{Linearized Gravity in Harmonic Gauge}

We consider perturbations $g_{\mu \nu} = \eta_{\mu \nu} + h_{\mu \nu} - \frac{1}{2}\eta_{\mu\nu}\eta^{\alpha\beta}h_{\alpha\beta}$ around a flat background metric
\begin{equation} \vspace*{-.1cm}
ds^2 = \eta_{\mu\nu} dx^{\mu} dx^{\nu} = -du^2 - 2dudr + 2r^2 \gamma_{z\bar{z}}dz d\bar{z}, \vspace*{-.1cm}
\end{equation}
and impose the harmonic gauge condition
\begin{equation}\label{eq:harga} \vspace*{-.1cm}
\nabla^{\mu}h_{\mu\nu}=0, \vspace*{-.1cm}
\end{equation} 
where $h_{\mu\nu}$ is the trace-reversed perturbation. In this gauge, the linearized Einstein equations are
\begin{equation} \vspace*{-.1cm}
\square h_{\mu\nu} = - 16 \pi G T_{\mu\nu} .
\end{equation}
Harmonic gauge leaves unfixed a set of residual diffeomorphisms $\xi$ that obey $\square \xi = 0$.  Coordinate expansions of the Einstein equations, harmonic gauge condition, and residual diffeomorphisms are in Appendix A.
 \vspace*{-.1cm}

\subsection{Boundary Conditions}

We choose falloffs of the matter stress tensor $T_{\mu \nu}$ consistent with a massless scalar field (see also~\cite{Campiglia2017, Pate2018}). This corresponds to 
\begin{equation} \label{eq:Gfalloff} \vspace*{-.1cm}
\begin{aligned}
G_{uu} \sim \mathcal{O}(r^{-2}), \ \ \ \ \ G_{ur} \sim \mathcal{O}(r^{-4}), \ \ \ \ \ G_{rr} \sim \mathcal{O}(r^{-4}), \\
G_{uA} \sim \mathcal{O}(r^{-2}), \ \ \ \ \ G_{rA} \sim \mathcal{O}(r^{-3}), \ \ \ \ \ G_{AB} \sim \mathcal{O}(r^{-1}). 
\end{aligned} \vspace*{-.1cm}
\end{equation}
These asymptotics for the stress tensor can be consistently captured by a metric with the following boundary behavior
 \begin{equation} \label{eq:hfalloff}
\begin{aligned} \vspace*{-.1cm}
h_{uu} \sim \mathcal{O}(r^{-1}\log r), \ \ \ \ \ h_{ur} \sim \mathcal{O}(r^{-1} \log r), \ \ \ \ \ h_{rr} \sim \mathcal{O}(r^{-1} \log r), \\
h_{uA} \sim \mathcal{O}(\log r), \ \ \ \ \ h_{rA} \sim \mathcal{O}(\log r), \ \ \ \ \ h_{AB} \sim \mathcal{O}(r \log r).
\end{aligned} \vspace*{-.1cm}
\end{equation}
Note that in harmonic gauge, logarithmic $r$-dependence is required for a consistent solution of the linearized Einstein equations with matter in four dimensions.  

We write a large-$r$ mode expansion and solve the Einstein equations and the harmonic gauge condition order-by-order in $r$.  These expansions are written out in Appendix A.  Throughout, we denote the term in the metric expansion with coefficient $\frac{1}{r^n}$ by the superscript $(n)$ and the term in the expansion with coefficient $\frac{\log r}{r^n}$ by a tilde with superscript $(n)$.  We will use the same notation for the modes of other fields in what follows. 

The residual diffeomorphisms for harmonic gauge are parameterized by the free data
\begin{equation}\label{eq:udata}
\{\xi^{u(1)}(u,z,\bz), \xi^{r(1)}(u,z,\bz), \xi^{A(2)}(u,z,\bz) \}.
\end{equation}
These are the modes of $\xi^\mu$ which can have arbitrary $u$-dependence, and solutions to  $\Box \xi=0$ can be found by recursively solving~(\ref{eq:vec1}-\ref{eq:vec2}) starting from these modes.  As detailed in Appendix B, we can use these arbitrary functions of $(u,z,\bz)$ in~(\ref{eq:udata}) to perform residual gauge fixing and arrive at the stronger falloffs
\begin{equation}
\begin{aligned}\label{eq:fixedfalloff}
  h_{uu} &= \sum_{n=2}^{\infty} \frac{h_{uu}^{(n)}}{r^n} + \sum_{n=1}^{\infty} \frac{ \tilde{h}_{uu}^{(n)} \log r}{r^n}, &h_{ur} = \sum_{n=2}^{\infty} \frac{h_{ur}^{(n)}}{r^n} + \sum_{n=2}^{\infty} \frac{ \tilde{h}_{ur}^{(n)} \log r}{r^n}, \\
  h_{rr} &= \sum_{n=3}^{\infty} \frac{h_{rr}^{(n)}}{r^n} + \sum_{n=3}^{\infty} \frac{ \tilde{h}_{rr}^{(n)} \log r}{r^n},  &h_{uA} = \sum_{n=1}^{\infty} \frac{h_{uA}^{(n)}}{r^n} + \sum_{n=1}^{\infty} \frac{ \tilde{h}_{uA}^{(n)} \log r}{r^n}, \\
  h_{rA} &= \sum_{n=1}^{\infty} \frac{h_{rA}^{(n)}}{r^n} + \sum_{n=2}^{\infty} \frac{ \tilde{h}_{rA}^{(n)} \log r}{r^n} , &h_{AB} = \sum_{n=-1}^{\infty} \frac{h_{AB}^{(n)}}{r^n} + \sum_{n=0}^{\infty} \frac{ \tilde{h}_{AB}^{(n)} \log r}{r^n}. 
\end{aligned}
\end{equation}

\subsection{Residual Symmetries}

We now consider the full set of residual diffeomorphisms that preserve the gauge-fixed falloffs~(\ref{eq:fixedfalloff}).  Since the arbitrary $u$ dependence of~(\ref{eq:udata}) has been removed by our residual gauge choice in Appendix B, we expect $\xi$ to be parameterized by functions of $(z,\bz)$.  In  Appendix C we show that the residual diffeomorphisms have the following large-$r$ behavior
\be\begin{array}{l}\label{eq:vect}
\xi^u=\frac{u}{2}D^AY_A+f+\mathcal{O}(r^{-2}\log r)\\
\xi^r=-\frac{r}{2}D^AY_A-\frac{u}{2}D^AY_A+\frac{1}{2}D^2f+Hr^{-1}+(\frac{u}{4}D^2[D^2+2]f-E)r^{-1}\log r+\mathcal{O}(r^{-2}\log r)\\
\xi^B=Y^B-D^B(\frac{u}{2}D^AY_A+f)r^{-1}+(\frac{u}{2}D^B[D^2+2]f+V^B)r^{-2}+\mathcal{O}(r^{-3}\log r) . 
\end{array}\ee
Here the free data are
\be\label{eq:alldiff} 
\{f(z,\bz), Y^z(z), H(z,\bz), E(z,\bz), V^A(z,\bz),...\}
\ee
with $Y^\bz(\bz)=\overline{Y^z}$ and the ellipsis denotes integration constants that appear at each subleading order when recursively solving~(\ref{eq:vec1}-\ref{eq:vec2}).\footnote{Since the residual diffeomorphisms are parameterized by $u$-independent functions, any further gauge fixing would only be able to fix certain metric components at one value of $u$, and will not further modify the large-$r$ falloffs in~(\ref{eq:fixedfalloff}).} 

The leading terms parameterized by $f(z,\bz)$ and $Y^z(z)$ correspond to supertranslations and superrotations, respectively.  These are the only modes of $\xi^\mu$ that contribute to a non-zero charge at null infinity~\cite{Wald:1999wa}, with linear terms given by~\cite{Barnich2011}
\be\label{eq:charge1}
\hat{Q}_\xi^{\mathcal{I}^+_-}= -\frac{1}{4\pi G}\int_{\mathcal{I}^+_-}d^2z\sqrt{\gamma}  \left[(f+\frac{1}{2}uD_AY^A)\gamma^{z\bar{z}}C_{u\bar{z}zr}^{(1)}-\frac{1}{2}Y^z C_{zrru}^{(3)}-\frac{1}{2}Y^\bz C_{\bz rru}^{(3)}\right],
\ee
where the leading modes of the Weyl tensor components are
\be\label{eq:weyl1}
C_{u\bar{z}zr}^{(1)} =\lim\limits_{r\rightarrow\infty} r C_{u\bar{z}zr}~~~C_{zrru}^{(3)}=\lim\limits_{r\rightarrow\infty} r^3 C_{zrru}.
\ee 
These correspond to the Weyl scalars $\Psi_2^0$ and $\Psi_1^0$, respectively, in the Newman-Penrose formalism~\cite{Newman1962}, up to a rescaling due to tetrad normalization for our celestial sphere metric.  For reference,~(\ref{eq:charge1}) is the linear part of (3.2) of~\cite{Barnich2011}  (see (4.4) of~\cite{Kapec2014} and (2.4) of~\cite{Pasterski2016} for expressions for the Weyl tensor components in terms of the mass and angular momentum aspects appearing in~\cite{Barnich2011}).  

The above residual diffeomorphism~(\ref{eq:vect}) produces the following inhomogeneous shifts in the leading modes of the chiral part of the sphere metric
\be\label{eq:shift}
\delta h_{zz}^{(-1)}=-uD_z^3Y^z-2D_z^2 f,~~~ \delta h_{zz}^{(0)}=u[D^2-2]D_z^2f+2D_zV_z .
\ee
Note that the $u$-independent early and late time behavior of $h_{zz}^{(0)}$ both shift under the residual diffeomorphism parameterized by $V_A(z,\bz)$.
As in the electromagnetic case~\cite{Himwich:2019dug}, the subleading soft theorem will correspond to a difference in the boundary values of this subleading-in-$r$ mode $h_{zz}^{(0)}$ of the sphere metric.  

\section{Conservation Law}
In what follows, we will work in units where $8\pi G=1$. 
Using the Einstein equations and the harmonic gauge condition, we find that the Weyl tensor modes in~(\ref{eq:weyl1}) evaluate to
\begin{equation}
  \begin{aligned}
    \gamma^{z\bar{z}}C_{u\bar{z}zr}^{(1)} &
    = \frac{1}{2}\tilde{h}_{uu}^{(1)} - \frac{1}{2} D^{\bar{z}}D^{\bar{z}} h_{\bar{z}\bar{z}}^{(-1)} + \frac{1}{6}\gamma^{z \bar{z}}T_{z\bar{z}}^{(1)}
    \end{aligned}
\end{equation}
and
\begin{equation}
  \partial_u C_{zrru}^{(3)} 
  = - T_{uz}^{(2)} - \frac{1}{2}D^{\bar{z}}T_{z\bar{z}}^{(1)} -\frac{1}{2} D_z \tilde{h}_{uu}^{(1)} + \frac{1}{2} D_z D^{\bar{z}}D^{\bar{z}} h_{\bar{z}\bar{z}}^{(-1)}.
\end{equation}
To investigate the superrotation charge in~(\ref{eq:charge1}), we consider the contribution at a fixed point on the celestial sphere\footnote{If we were not restricted to CKVs this would amount to setting $Y^z\rightarrow\delta^2(z-w), Y^\bz\rightarrow0, f\rightarrow0$.  Equating this to~(\ref{eq:charge1}) requires the boundary condition~(\ref{eq:vaccend}).}
\begin{equation}
  \begin{aligned}
    \int du \partial_u ( C_{zrru}^{(3)} + u D^\bz C_{u\bar{z}zr}^{(1)}) = &- \int du T_{uz}^{(2)}  + \frac{1}{2}  \int du D_z u \partial_u \tilde{h}_{uu}^{(1)} - \frac{1}{2} \int du D_z D^{\bar{z}}D^{\bar{z}} u \partial_u h_{\bar{z}\bar{z}}^{(-1)} \\
    &- \frac{1}{2} \int du D^{\bar{z}}T_{z\bar{z}}^{(1)} + \frac{1}{6} \int du  D^{\bar{z}}\partial_u(uT_{z\bar{z}}^{(1)}).
  \end{aligned}
\end{equation}
Stress tensor conservation gives
\begin{equation}
  \partial_u T_{rr}^{(4)} = - \gamma^{AB}T_{AB}^{(1)},
\end{equation}
and since $T_{rr}$ vanishes at the boundaries of $\mathcal{I}^+$, we find that the $u$-integral of the trace of $T_{AB}^{(1)}$ vanishes.  
Using $T_{uu}^{(2)} = \partial_u \tilde{h}_{uu}^{(1)}$ and evaluating the boundary terms gives
\begin{equation}
  \begin{aligned} \label{eq:Mcharge}
    \big(  C_{zrru}^{(3)} + u D^\bz C_{u\bar{z}zr}^{(1)} \big) \Big{|}^{\mathcal{I}^+_+}_{\mathcal{I}^+_-}  = &- \int du \big(T_{uz}^{(2)} - \frac{1}{2} u D_z T_{uu}^{(1)} \big) - \frac{1}{2} \int du D_z D^{\bar{z}}D^{\bar{z}} u \partial_u h_{\bar{z}\bar{z}}^{(-1)} .
  \end{aligned}
\end{equation}
The Ward identity for the linearized superrotation charge~(\ref{eq:charge1}), which contains a convolution of the left-hand side of~(\ref{eq:Mcharge}) with a CKV $Y^z$, was demonstrated in~\cite{Kapec2014} using the subleading soft theorem.  When this CKV is specified to the particular complexified form $Y^z=\frac{1}{w-z},~Y^\bz=0$, the contribution from the second term on the right-hand side of~(\ref{eq:Mcharge}) is proportional to the 2D stress tensor for 4D gravity~\cite{Kapec2017}.  In ${\cal S}$-matrix elements, this term inserts a subleading soft graviton.    This subleading soft graviton mode also appeared in~\cite{Pasterski2016} as the spin memory observable.  

So far, we have only performed computations near $\mathcal{I}^+$ but an analogous story holds near $\mathcal{I}^-$, and the additional input of a matching condition and falloffs at the boundaries of null infinity is required to make statements about symmetries of the $\mathcal{S}$-matrix.
 The relevant matching conditions for the Weyl tensor components are given in (2.12)-(2.13) of~\cite{Pasterski2016}
 \be\label{eq:matching}
C_{u\bar{z}zr}^{(1)}\Big{|}_{\mathcal{I}^+_-}=C_{v\bar{z}zr}^{(1)}\Big{|}_{\mathcal{I}^-_+}~,~~~~\p_{[z}C_{\bz] rru}^{(3)} \Big{|}_{\mathcal{I}^+_-}=\p_{[z}C_{\bz] rrv}^{(3)} \Big{|}_{\mathcal{I}^-_+}.
\ee
The analyses~\cite{Kapec2014,Pasterski2016} looked in particular at spacetimes that start and end in vacuum with massless matter that enters and exits through past and future null infinity.  This amounts to setting
\be\label{eq:vaccend}
   \big(  C_{zrru}^{(3)} + u D^\bz C_{u\bar{z}zr}^{(1)} \big) \Big{|}_{\mathcal{I}^+_+} =   \big( C_{zrrv}^{(3)} + v D^\bz C_{v\bar{z}zr}^{(1)} \big) \Big{|}_{\mathcal{I}^-_-}=0. 
\ee
The matching of the Weyl scalars was used in~\cite{Kapec2014} to recast~(\ref{eq:Mcharge}) and its past null infinity counterpart as a conservation law.   In the following section we recast the charge in terms of a difference in the boundary values of $h_{zz}^{(0)}$.  Then, in section~\ref{sec:spinmem}, we recast the spin memory observable in the same terms.

\subsection{Expression as a Boundary Difference}
We now rewrite the subleading soft graviton mode in terms of a change in the boundary values of asymptotic data using the Einstein equations in harmonic gauge, which give
\begin{equation}
  - 2 T_{\bar{z}\bar{z}}^{(1)} = [\square h_{\bar{z}\bar{z}}]^{(1)} = 2 \partial_u h_{\bar{z}\bar{z}}^{(0)} + [D^2 - 2] h_{\bar{z}\bar{z}}^{(-1)} .
\end{equation}
Recall from the previous subsection that stress tensor conservation implies that the $u$-integral of $D^{{z}}T_{z\bar{z}}^{(1)}$ vanishes. We note that stress tensor conservation also gives
\begin{equation}
  \partial_uT_{r\bz}^{(3)} = D^BT_{B\bz}^{(1)}.
\end{equation}
Then, by taking $T_{r\bz}^{(3)}$ to vanish at the boundaries of $\mathcal{I}^+$, the $u$-integral of $D^{\bz}T_{\bz\bz}^{(1)}$ vanishes as well.  Using also that $D^{\bz}T_{\bz\bz}^{(1)}$ falls off faster than $u^{-1}$, we have
\be
\int du~u\p_u D^\bz [D^2 - 2]h_{\bar{z}\bar{z}}^{(-1)} =-2\int du~u\p_u^2 D^\bz  h_{\bar{z}\bar{z}}^{(0)} .
\ee
A straightforward computation gives
\begin{equation}
\begin{aligned}
 [D^2 +1] D_zD^{\bar{z}}D^{\bar{z}} h_{\bar{z}\bar{z}}^{(-1)} &=  D_zD^{\bar{z}}D^{\bar{z}} [D^2 - 2] h_{\bar{z}\bar{z}}^{(-1)}  .
  \end{aligned}
\end{equation}
With this we can rewrite
\begin{equation}\label{eq:shift1}
  \begin{aligned}
    \scalemath{0.9}{\big(  C_{zrru}^{(3)}+ u D^{\bar{z}}C_{u\bar{z}zr}^{(1)} \big) \Big{|}^{\mathcal{I}^+_+}_{\mathcal{I}^+_-}}  = &\scalemath{0.85}{- \int du \big(T_{uz}^{(2)} - \frac{1}{2} u D_z T_{uu}^{(1)} \big) - [D^2 + 1]^{-1} D_zD^{\bar{z}}D^{\bar{z}}(1 - u \partial_u) h_{\bar{z}\bar{z}}^{(0)} \Big{|}^{\mathcal{I}^+_+}_{\mathcal{I}^+_-}}.
  \end{aligned}
\end{equation}
Note the appearance of the operator $(1 - u \partial_u)$, as in electromagnetism~\cite{Himwich:2019dug}. This subtracts off the linear $u$-growth in $h_{\bz\bz}^{(0)}$. At early and late times, the matter stress tensor vanishes and using~(\ref{eq:shift}) we can describe the asymptotic behavior of the metric perturbations near $\mathcal{I}^{+}_{\pm}$ as
\begin{equation}\label{eq:latetime}
  h_{\bz\bz,\pm}^{(-1)} =-uD_\bz^3\hat{Y}^\bz(\bz) -2 D_\bz^2 \hat{f}^{\pm}(z,\bar{z}), \ \ \ h_{\bz\bz,\pm}^{(0)} = u [D^2 - 2] D_\bz^2 \hat{f}^{\pm}(z,\bar{z}) + 2D_\bz \hat{V}_\bz^{\pm}(z,\bar{z}).
\end{equation}
The notation is intended to reflect that used for the residual vector field $V^A$ and supertranslation Goldstone mode $f(z,\bar{z})$ of ~\cite{Strominger2016a} (denoted $C(z,\bz)$ there, see also~\cite{He:2014laa}), with the carat emphasizing the distinction that such a diffeomorphism would shift both the $\mathcal{I}^+_+$ and $\mathcal{I}^+_-$ values of the respective quantities but would not affect the difference between their boundary values.  We have also allowed for a superrotation parameterized by $\hat{Y}^\bz(\bz)$, which is in the kernel of the differential operators acting on $h_{\bz\bz}^{(-1)}$ in~(\ref{eq:Mcharge}) and so will not affect the conclusions that follow regarding the memory effect.\footnote{As long as we consider asymptotically flat solutions without snapping cosmic strings~\cite{Strominger:2016wns}, there will be no transition between differently superrotated vacua (hence we drop a $\pm$ superscript for $\hat{Y}^\bz$).}   We thus have  
\begin{equation}\label{eq:shift2}
  \begin{aligned}
    D_zD^{\bar{z}}D^{\bar{z}}(1 - u \partial_u)h_{\bar{z}\bar{z}}^{(0)}\Big{|}^{\mathcal{I}^+_+}_{\mathcal{I}^+_-} &= [D^2 + 1]D_zD_z\hat{V}^z\Big{|}^{\mathcal{I}^+_+}_{\mathcal{I}^+_-},
    \end{aligned}
\end{equation}
which finally gives 
\begin{equation} \label{eq:Vcharge}
  \begin{aligned}
    \big(  C_{zrru}^{(3)} + u D^\bz C_{u\bar{z}zr}^{(1)} \big) \Big{|}^{\mathcal{I}^+_+}_{\mathcal{I}^+_-}  = &- \int du \big(T_{uz}^{(2)} - \frac{1}{2} u D_z T_{uu}^{(2)} \big) - D_zD_z\hat{V}^z\Big{|}^{\mathcal{I}^+_+}_{\mathcal{I}^+_-}.
  \end{aligned}
\end{equation}
We have rewritten the soft part of the superrotation charge as a difference in the boundary values of in $\hat{V}_A$.  As in the discussion following~(\ref{eq:Mcharge}), one can also use the matching~(\ref{eq:matching}) and boundary conditions~(\ref{eq:vaccend}) to recast the difference in $\hat{V}_A$ in terms of stress tensor fluxes.  The soft part of the charge, given in (5.13) of~\cite{Kapec2014}, is
\be
Q_S^+(Y^z,Y^\bz=0)=\frac{1}{2}\int_{\mathcal{I}^+} \sqrt{\gamma}d^2z du Y^z D_z D^\bz D^\bz u\p_u h_{\bz\bz}^{(-1)}=\int \sqrt{\gamma}d^2z Y^z D_zD_z\hat{V}^z\Big{|}^{\mathcal{I}^+_+}_{\mathcal{I}^+_-},
\ee
where we have complexified the superrotations. In particular we find
\be
T^{CFT}_{ww}=2i Q_{S}^+(Y^z=\frac{1}{w-z},Y^\bz=0)
\ee
as mentioned above. This soft charge generates an inhomogeneous shift in the News tensor ($\p_u h_{zz}^{(-1)}$ in our notation)
\be\label{eq:chargenews}
[Q_S^+,h^{(-1)}_{zz}]=i u D_z^3Y^z.
\ee
Acting on the vacuum, the soft charge inserts a soft graviton rather than leaving it invariant, providing a notion of Goldstone bosons within the context of asymptotic symmetries and soft theorems (see~\cite{Strominger:2013jfa}). For the supertranslation case, \cite{He:2014laa} introduced a symplectic pairing between the Goldstone mode and a conjugate soft mode, which~\cite{Donnay:2018neh} cast in the conformal basis~\cite{Pasterski:2017kqt}.~~\cite{Donnay:2018neh} also proposed the superrotation analog of the Goldstone mode, whose shift is parameterized by $Y^A$.  From~(\ref{eq:chargenews}) we see that $\Delta \hat{V}_A$ is related to the conjugate of the  Goldstone mode.  

In the supertranslation case, the symplectically paired modes are $C(z,\bz)$ and $N(z,\bz)$ of~\cite{He:2014laa}, where $C(z,\bz)$ parameterizes the supertranslation Goldstone mode and $N(z,\bz)$ parameterizes the difference in boundary values.  Both are at radiative order. Here, in the superrotation case, the difference in boundary values of $h_{zz}^{(0)}$ and constant-in-$u$ Goldstone mode $\p_u h_{zz}^{(-1)}$ are separated by an order of $r$.  We leave the detailed study of the symplectic pairing to future work.\footnote{ Note that here the $\Delta \hat{V}_A$ that appears in our recasting of the soft graviton mode has a priori no restrictions, while the superrotated vacua are parameterized only by holomorphic $Y^z(z)$. It suggests that a thorough analysis of the appropriate symplectic pairing will connect to an ongoing question in the literature of whether superrotations should be enhanced to $\mathrm{Diff}(S^2)$~\cite{Campiglia2014} (see also~\cite{Conde:2016rom} for an alternate proposal).  This involves a modification of the boundary falloffs but allows one to invert the soft theorem from the Ward identity.  On the other hand there may be a more natural way of projecting onto the part of $\Delta \hat{V}_A$ that provides the natural symplectic partner to the superrotation Goldstone mode, which we hope to address in future work.}

\subsection{Spin Memory}\label{sec:spinmem}

In~\cite{Pasterski2016}, the spin memory observable was defined to be an accumulated time delay $\Delta^+u$ between two counter-propagating light beams for a BMS detector arranged in a ring with circumference $2\pi L$
\be\label{eq:spindelay}
\Delta^+u=\frac{1}{2\pi L}\int du \oint_{\mathcal{C}}(D^z h_{zz}^{(-1)}dz+D^\bz h_{\bz\bz}^{(-1)}d\bz).
\ee
By Stokes's theorem, this is proportional to a surface integral of the curl $\mathrm{Im}[D_z^2 h^{zz(-1)}]$ over the region bounded by $\mathcal{C}$.  This curl has the nice feature of projecting out the linearly growing piece in the radiative metric~(\ref{eq:latetime}).  The expression for $\Delta^+u$ was shown to be equal to
\be
\Delta^+ u=-\frac{1}{\pi^2L}\mathrm{Im} \int_{D_\mathcal{C}}d^2w \gamma_{w\bw}\int d^2 z \p_\bz \mathcal{G}(z;w)\left[C_{zrru}^{(3)}\Big{|}^{\mathcal{I}^+_+}_{\mathcal{I}^+_-}+\int_{\mathcal{I}^+} du T_{uz}^{(2)}\right]\vspace*{.3cm}
\ee
where we have introduced the Green's function~\cite{Pasterski2016}
\vspace*{-.2cm}\be
\mathcal{G}(z;w)=\log \sin^2\frac{\Theta}{2},~~\sin^2\frac{\Theta(z,w)}{2}\equiv\frac{|z-w|^2}{(1+w\bw)(1+z\bz)} \vspace*{-.2cm}
\ee
which obeys \vspace*{-.2cm}
\be\label{eq:source}
\scalemath{1}{\p_z\p_\bz \mathcal{G}(z;w)=2\pi\delta^2(z-w)-\frac{1}{2}\gamma_{z\bz}} .\vspace*{-.2cm}
\ee 
Now from~(\ref{eq:Vcharge}), we have
\begin{equation} 
  \begin{aligned}
  \int_{\mathcal{I}^+} du ~2 D_{[\bz} T_{z]u}^{(2)} =-2D_{[\bz}C_{z]rru}^{(3)}-D_\bz D_z D_z \hat{V}^z+D_z D_\bz D_\bz \hat{V}^\bz\Big{|}^{\mathcal{I}^+_+}_{\mathcal{I}^+_-},
   \end{aligned}
\end{equation}
where the curl projects out the $\int u D_z T_{uu}^{(2)}$ in~(\ref{eq:Vcharge}).
We thus have
\be\begin{aligned}\label{eq:delusimp}
\Delta^+ u&=\frac{1}{\pi^2L}\mathrm{Im} \int_{D_\mathcal{C}}d^2w \gamma_{w\bw}\int d^2 z\p_z\p_\bz \mathcal{G}(z;w)D_z\hat{V}^z\Big{|}^{\mathcal{I}^+_+}_{\mathcal{I}^+_-}\\
&=\frac{2}{\pi L}\mathrm{Im} \int_{D_\mathcal{C}}d^2w \gamma_{w\bw}\left[D_w \hat{V}^w-\frac{1}{4\pi}\int d^2 z \gamma_{z\bz} D_z\hat{V}^z\right]\Big{|}^{\mathcal{I}^+_+}_{\mathcal{I}^+_-}\\
&=\frac{1}{i\pi L} \int_{D_\mathcal{C}}d^2w [D_w \hat{V}_\bw-D_\bw \hat{V}_w]\Big{|}^{\mathcal{I}^+_+}_{\mathcal{I}^+_-} \\
\end{aligned}
\ee
using the fact that the integral of a curl over the full $z$-sphere vanishes to kill the contribution from the second term in~(\ref{eq:source}).  Using
\be
d^2w=dx\wedge dy=\frac{i}{2}dw\wedge d\bw, 
\ee
(\ref{eq:delusimp}) is beautifully recast as
\be\label{eq:spinresult}
\Delta^+ u=\frac{1}{2\pi L}  \oint_{\mathcal{C}} \hat{V}_A dx^A \Big{|}^{\mathcal{I}^+_+}_{\mathcal{I}^+_-} .
\ee
We learn that spin memory measures the change between early and late time values of the the contour integral of the subleading soft mode $\hat{V}^A$ that we have identified in this note.\footnote{As in (5.9) of~\cite{Pasterski2016}, we could consider spacetimes that satisfy~(\ref{eq:vaccend}), and combine contributions from past and future null infinity $\Delta \tau\equiv \Delta^+ u-\Delta^- v$, so as to cancel the Weyl tensor contribution, and write $\Delta \tau$ as in terms of fluxes of $T_{uz}$ and $T_{vz}$.  However, this additional restriction is not needed to equate the observable time delay identified in~\cite{Pasterski2016} to a difference in the boundary values of $\hat{V}_A$ via~(\ref{eq:spinresult}). } \vspace*{.5cm}

\section*{Acknowledgements}

We are grateful to Dan Kapec, Monica Pate, and Ana Raclariu for useful discussions, and in particular Andrew Strominger and Burkhard Schwab for collaboration at an early stage of this project. E.H. is funded by the National Science Foundation through a Graduate Research Fellowship under grant DGE-1745303. Z.M. is supported in part by the Erwin Schr\"odinger JRF fund.
S.P. is supported by the National Science Foundation through a Graduate Research Fellowship under grant DGE-1144152 and by the Hertz Foundation through a Harold and Ruth Newman Fellowship.

\appendix

\section{Asymptotic Expansions}

In components, the linearized Einstein equations $\square h_{\mu\nu} = - 16 \pi G T_{\mu\nu}$ have right-hand side
\begin{equation}\scalemath{0.9}{
\begin{aligned}
\square h_{uu} &= \left( \partial_r^2 - 2 \partial_r \partial_u - \frac{2}{r}(\partial_u - \partial_r) + \frac{1}{r^2}D^2\right) h_{uu} \\
\square h_{ur} &= \left( \partial_r^2 - 2 \partial_r \partial_u - \frac{2}{r}(\partial_u - \partial_r) + \frac{1}{r^2}D^2\right) h_{ur} + \frac{2}{r^2}(h_{uu} - h_{ur})  - \frac{2}{r^3}D^A h_{uA}\\
\square h_{rr} &= \left( \partial_r^2 - 2 \partial_r \partial_u - \frac{2}{r}(\partial_u - \partial_r) + \frac{1}{r^2}D^2\right) h_{rr} - \frac{4}{r^3}D^A h_{Ar} + \frac{4}{r^2}(h_{ur} - h_{rr}) + \frac{2}{r^4}\gamma^{CB}h_{CB} \\
\square h_{uA} &= \left( \partial_r^2 - 2 \partial_r \partial_u  + \frac{1}{r^2}D^2\right) h_{uA} - \frac{1}{r^2} h_{uA} - \frac{2}{r}\partial_A(h_{uu} - h_{ur}) \\
\square h_{rA} &= \left( \partial_r^2 - 2 \partial_r \partial_u  + \frac{1}{r^2}D^2\right) h_{rA} - \frac{5}{r^2} h_{rA} + \frac{4}{r^2}h_{uA}- \frac{2}{r}\partial_A(h_{ur} - h_{rr}) - \frac{2}{r^3}D^C h_{CA} \\
\square h_{AB} &= \left( \partial_r^2 - 2 \partial_r \partial_u + \frac{2}{r}(\partial_u - \partial_r) + \frac{1}{r^2}D^2\right) h_{AB} \\
&- \frac{2}{r}D_A(h_{uB} - h_{rB}) - \frac{2}{r}D_B(h_{uA} - h_{rA}) + 2 \gamma_{AB}(h_{uu} - 2h_{ur} + h_{rr}). \\
\end{aligned}}
\end{equation}
We expand the components of the harmonic gauge condition $\nabla^{\mu}h_{\mu \nu} =0$ as
\begin{equation}\scalemath{0.9}{
\begin{aligned}
\nabla^{\mu}h_{\mu u} &= - \partial_u h_{ur} - \partial_r(h_{uu} - h_{ur}) - \frac{2}{r}(h_{uu} - h_{ur}) + \frac{1}{r^2}D^Ah_{uA} \\
\nabla^{\mu}h_{\mu r} &= - \partial_u h_{rr} - \partial_r(h_{ur} - h_{rr}) - \frac{2}{r}(h_{ur} - h_{rr}) + \frac{1}{r^2}D^Ah_{rA} - \frac{1}{r^3}\gamma^{AB}h_{AB} \\
\nabla^{\mu}h_{\mu A} &= - \partial_u h_{rA} - \partial_r(h_{uA} - h_{rA}) - \frac{2}{r}(h_{uA} - h_{rA}) + \frac{1}{r^2}D^Bh_{AB} .
\end{aligned}}
\end{equation}
The residual diffeomorphisms $\xi^{\mu}$ that preserve the harmonic gauge condition obey $\square \xi^{\mu} = 0$, which is
\begin{equation}\scalemath{0.9}{
\begin{aligned}
\left(\partial_r^2 - 2\partial_r\partial_u - \frac{2}{r}(\partial_u - \partial_r) + \frac{1}{r^2}D^2\right)\xi_u  &= 0 \\
\left(\partial_r^2 - 2\partial_r\partial_u - \frac{2}{r}(\partial_u - \partial_r) + \frac{1}{r^2}D^2\right)\xi_r - \frac{2}{r^3}D^A\xi_A + \frac{2}{r^2}(\xi_u - \xi_r)  &= 0 \\
\left(\partial_r^2 - 2\partial_r\partial_u  + \frac{1}{r^2}D^2\right)\xi_A - \frac{1}{r^2}\xi_A - \frac{2}{r}\partial_A(\xi_u - \xi_r) &=0 .
\end{aligned}}
\end{equation}
As noted above, logarithmic-in-$r$ modes are required for a consistent solution of the linearized Einstein equations with matter in four dimensions. 
In terms of the modes that appear in~(\ref{eq:fixedfalloff}), the asymptotic expansion of the Einstein equations is 
\begin{equation} \label{eq:nEinstein}\scalemath{0.9}{
\begin{aligned}
[\square h_{uu}]^{(n)} &= 2(n-2) \partial_uh_{uu}^{(n-1)} + [D^2 + (n-2)(n-3)]h_{uu}^{(n-2)}\\
&+ (5-2n) \tilde{h}_{uu}^{(n-2)} - 2\partial_u\tilde{h}_{uu}^{(n-1)} \\
[\square h_{ur}]^{(n)} &= 2(n-2) \partial_uh_{ur}^{(n-1)} + [D^2 + (n-2)(n-3) - 2]h_{ur}^{(n-2)}+ 2h_{uu}^{(n-2)}- 2D^Ah_{uA}^{(n-3)} \\
&+ (5-2n) \tilde{h}_{ur}^{(n-2)} - 2\partial_u\tilde{h}_{ur}^{(n-1)} \\
[\square h_{rr}]^{(n)} &= 2(n-2) \partial_uh_{rr}^{(n-1)} + [D^2 + (n-2)(n-3)]h_{rr}^{(n-2)}\\
&+ 4\left(h_{ur}^{(n-2)}- h_{rr}^{(n-2)}\right) - 4D^Ah_{rA}^{(n-3)} + 2\gamma^{AB}h_{AB}^{(n-4)} + (5-2n) \tilde{h}_{rr}^{(n-2)} - 2\partial_u\tilde{h}_{rr}^{(n-1)} \\
[\square h_{uA}]^{(n-1)} &= 2(n-2) \partial_uh_{uA}^{(n-2)} + [D^2 + (n-3)(n-2) - 1]h_{uA}^{(n-3)}- 2\partial_A\left(h_{uu}^{(n-2)} - h_{ur}^{(n-2)}\right) \\
&+ (5-2n) \tilde{h}_{uA}^{(n-3)} - 2\partial_u\tilde{h}_{uA}^{(n-2)} \\
[\square h_{rA}]^{(n-1)} &= 2(n-2) \partial_uh_{rA}^{(n-2)} + [D^2 + (n-3)(n-2) - 1]h_{rA}^{(n-3)}- 2\partial_A\left(h_{ur}^{(n-2)} - h_{rr}^{(n-2)}\right) \\
&- 2D^Bh_{AB}^{(n-4)} + 4(h_{uA}^{(n-3)}-h_{rA}^{(n-3)})+ (5-2n) \tilde{h}_{rA}^{(n-3)} - 2\partial_u\tilde{h}_{rA}^{(n-2)} \\
[\square h_{AB}]^{(n-2)} &= 2(n-2) \partial_uh_{AB}^{(n-3)} + [D^2 + (n-2)(n-3) - 2]h_{AB}^{(n-4)}\\
&- 2\left(D_Ah_{uB}^{(n-3)} - D_Ah_{rB}^{(n-3)} + D_Bh_{uA}^{(n-3)} - D_Bh_{rA}^{(n-3)}\right) \\
&~~~+ 2 \gamma_{AB}\left(h_{uu}^{(n-2)} - 2h_{ur}^{(n-2)} + h_{rr}^{(n-2)} \right) + (5-2n)\tilde{h}_{AB}^{(n-4)} - 2 \partial_u\tilde{h}_{AB}^{(n-3)}  . 
\end{aligned}}
\end{equation}
The expansion for the Einstein equations with log coefficients is
\begin{equation}\label{eq:nEinsteinLog}\scalemath{0.9}{
\begin{aligned}
[\square \tilde{h}_{uu}]^{(n)} &= 2(n-2) \partial_u\tilde{h}_{uu}^{(n-1)} + [D^2 + (n-2)(n-3)] \tilde{h}_{uu}^{(n-2)} \\
[\square \tilde{h}_{ur}]^{(n)} &= 2(n-2) \partial_u\tilde{h}_{ur}^{(n-1)} + [D^2 + (n-2)(n-3) - 2]\tilde{h}_{ur}^{(n-2)} + 2\tilde{h}_{uu}^{(n-2)} - 2D^A\tilde{h}_{uA}^{(n-3)} \\
[\square \tilde{h}_{rr}]^{(n)} &= 2(n-2) \partial_u\tilde{h}_{rr}^{(n-1)} + [D^2 + (n-2)(n-3)]\tilde{h}_{rr}^{(n-2)} + 4\left(\tilde{h}_{ur}^{(n-2)} - \tilde{h}_{rr}^{(n-2)}\right) \\
&~~~- 4D^A\tilde{h}_{rA}^{(n-3)} + 2\gamma^{AB}\tilde{h}_{AB}^{(n-4)} \\
[\square \tilde{h}_{uA}]^{(n-1)} &= 2(n-2) \partial_u\tilde{h}_{uA}^{(n-2)} + [D^2 + (n-3)(n-2) - 1]\tilde{h}_{uA}^{(n-3)} - 2\partial_A\left(\tilde{h}_{uu}^{(n-2)} - \tilde{h}_{ur}^{(n-2)}\right) \\
[\square \tilde{h}_{rA}]^{(n-1)} &= 2(n-2) \partial_u\tilde{h}_{rA}^{(n-2)} + \left[D^2 + (n-3)(n-2) - 1\right]\tilde{h}_{rA}^{(n-3)} - 2\partial_A\left(\tilde{h}_{ur}^{(n-2)} - \tilde{h}_{rr}^{(n-2)}\right) \\
&~~~- 2D^B\tilde{h}_{AB}^{(n-4)} + 4(\tilde{h}_{uA}^{(n-3)}-\tilde{h}_{rA}^{(n-3)}) \\
[\square \tilde{h}_{AB}]^{(n-2)} &= 2(n-2) \partial_u\tilde{h}_{AB}^{(n-3)} + [D^2 + (n-2)(n-3) - 2)]\tilde{h}_{AB}^{(n-4)} \\
&~~~- 2\left(D_A\tilde{h}_{uB}^{(n-3)} - D_A\tilde{h}_{rB}^{(n-3)} + D_B\tilde{h}_{uA}^{(n-3)} - D_B\tilde{h}_{rA}^{(n-3)}\right) \\
&~~~+ 2 \gamma_{AB}\left(\tilde{h}_{uu}^{(n-2)} - 2\tilde{h}_{ur}^{(n-2)} + \tilde{h}_{rr}^{(n-2)} \right)  . 
\end{aligned}}
\end{equation}
We can also expand the harmonic gauge conditions
\begin{equation}\scalemath{0.9}{
\begin{aligned}\label{eq:hg1}
[\nabla^{\mu}h_{\mu u}]^{(n)} &= - \partial_u h_{ur}^{(n)} + (n-3)\left(h_{uu}^{(n-1)} - h_{ur}^{(n-1)}\right) + D^Ah_{uA}^{(n-2)}\\
&~~~- \left(\tilde{h}_{uu}^{(n-1)} - \tilde{h}_{ur}^{(n-1)}\right) \\
[\nabla^{\mu}h_{\mu r}]^{(n)} &= - \partial_u h_{rr}^{(n)} + (n-3)\left(h_{ur}^{(n-1)} - h_{rr}^{(n-1)}\right) + D^Ah_{rA}^{(n-2)}- \gamma^{AB}h_{AB}^{(n-3)} \\
&~~~- \left(\tilde{h}_{ur}^{(n-1)} - \tilde{h}_{rr}^{(n-1)}\right) \\
[\nabla^{\mu}h_{\mu A}]^{(n-1)} &= - \partial_u h_{rA}^{(n-1)} + (n-4)\left(h_{uA}^{(n-2)} - h_{rA}^{(n-2)}\right) + D^Bh_{BA}^{(n-3)}\\
&~~~- \left(\tilde{h}_{uA}^{(n-2)} - \tilde{h}_{rA}^{(n-2)}\right)  .
\end{aligned}}
\end{equation}
The harmonic gauge condition at logarithmic order is
\begin{equation}\scalemath{0.9}{
\begin{aligned}\label{eq:hg2}
[\nabla^{\mu}\tilde{h}_{\mu u}]^{(n)} &= - \partial_u \tilde{h}_{ur}^{(n)} + (n-3)\left(\tilde{h}_{uu}^{(n-1)} - \tilde{h}_{ur}^{(n-1)}\right) + D^A\tilde{h}_{uA}^{(n-2)} \\
[\nabla^{\mu}\tilde{h}_{\mu r}]^{(n)} &= - \partial_u \tilde{h}_{rr}^{(n)} + (n-3)\left(\tilde{h}_{ur}^{(n-1)} - \tilde{h}_{rr}^{(n-1)}\right) + D^A\tilde{h}_{rA}^{(n-2)} - \gamma^{AB}\tilde{h}_{AB}^{(n-3)} \\
[\nabla^{\mu}\tilde{h}_{\mu A}]^{(n-1)} &= - \partial_u \tilde{h}_{rA}^{(n-1)} + (n-4)\left(\tilde{h}_{uA}^{(n-2)} - \tilde{h}_{rA}^{(n-2)}\right) + D^B\tilde{h}_{BA}^{(n-3)} . 
\end{aligned}}
\end{equation}
The expansion of the harmonic gauge conditions on residual diffeomorphisms is 
\begin{equation}\label{eq:vec1}\scalemath{0.9}{
\begin{aligned}
[\square \xi_u]^{(n)} &= 2(n-2)\partial_u \xi_u^{(n-1)} + \left[D^2 +(n-2)(n-3) \right]\xi_u^{(n-2)}\\
&~~~+ (5-2n)\tilde{\xi}_u^{(n-2)} - 2\partial_u\tilde{\xi}_u^{(n-1)} \\
[\square \xi_r]^{(n)} &= 2(n-2)\partial_u \xi_r^{(n-1)} + \left[D^2 +(n-2)(n-3) - 2\right]\xi_r^{(n-2)}+ 2 \xi_u^{(n-2)}- 2 D^A\xi_A^{(n-3)} \\
&~~~+ (5-2n)\tilde{\xi}_r^{(n-2)} - 2\partial_u\tilde{\xi}_r^{(n-1)} \\
[\square \xi_A]^{(n-1)} &= 2(n-2)\partial_u \xi_A^{(n-2)} + \left[D^2 +(n-2)(n-3) -1 \right]\xi_A^{(n-3)}- 2 \partial_A\left(\xi_u^{(n-2)} - \xi_r^{(n-2)}\right) \\
&~~~+ (5-2n)\tilde{\xi}_A^{(n-3)} - 2\partial_u\tilde{\xi}_A^{(n-2)} ,
\end{aligned}}
\end{equation}
and at logarithmic order is
\begin{equation}\label{eq:vec2}\scalemath{0.9}{
\begin{aligned}
[\square \tilde{\xi}_u]^{(n)} &= 2(n-2)\partial_u \tilde{\xi}_u^{(n-1)} + \left[D^2 +(n-2)(n-3) \right]\tilde{\xi}_u^{(n-2)} \\
[\square \tilde{\xi}_r]^{(n)} &= 2(n-2)\partial_u \tilde{\xi}_r^{(n-1)} + \left[D^2 +(n-2)(n-3) - 2\right]\tilde{\xi}_r^{(n-2)} + 2 \tilde{\xi}_u^{(n-2)} - 2 D^A\tilde{\xi}_A^{(n-3)} \\
[\square \tilde{\xi}_A]^{(n-1)} &= 2(n-2)\partial_u \tilde{\xi}_A^{(n-2)} + \left[D^2 +(n-2)(n-3) -1 \right]\tilde{\xi}_A^{(n-3)} - 2 \partial_A\left(\tilde{\xi}_u^{(n-2)} - \tilde{\xi}_r^{(n-2)}\right) .
\end{aligned}}
\end{equation}
Under such a diffeomorphism, the flat background metric components shift as
\be\begin{array}{rl}\label{eq:shiftpower}
\left[\delta g_{uu}\right]^{(n)}&=-2\p_u\xi^{u(n)}-2\p_u\xi^{r(n)}\\
\left[\delta g_{ur}\right]^{(n)}&=-\p_u\xi^{u(n)}+(n-1)\xi^{u(n-1)}+(n-1)\xi^{r(n-1)}-\tilde{\xi}^{u(n-1)}-\tilde{\xi}^{r(n-1)}\\
\left[\delta g_{rr}\right]^{(n)}&=2(n-1)\xi^{u(n-1)}-2\tilde{\xi}^{u(n-1)}\\
\left[\delta g_{uA}\right]^{(n-1)}&=\gamma_{AB}\p_u\xi^{B(n+1)}-\p_A\xi^{u(n-1)}-\p_A\xi^{r(n-1)}\\
\left[\delta g_{rA}\right]^{(n-1)}&=-\p_A\xi^{u(n-1)}-n\gamma_{AB}\xi^{B(n)}+\gamma_{AB}\tilde{\xi}^{B(n)}\\
\left[\delta g_{AB}\right]^{(n-2)}&=\gamma_{BC}D_A\xi^{C(n)}+\gamma_{AC} D_B\xi^{C(n)}+2\gamma_{AB}\xi^{r(n-1)} ,
\end{array}\ee
and at logarithmic order as
\be\begin{array}{rl}\label{eq:shiftlog}
\left[\delta \tilde{g}_{uu}\right]^{(n)}&=-2\p_u\tilde{\xi}^{u(n)}-2\p_u \tilde{\xi}^{r(n)}\\
\left[\delta \tilde{g}_{ur}\right]^{(n)}&=-\p_u \tilde{\xi}^{u(n)}+(n-1)\tilde{\xi}^{u(n-1)}+(n-1)\tilde{\xi}^{r(n-1)}\\
\left[\delta\tilde{g}_{rr}\right]^{(n)}&=2(n-1)\tilde{\xi}^{u(n-1)}\\
\left[\delta \tilde{g}_{uA}\right]^{(n-1)}&=\gamma_{AB}\p_u\tilde{\xi}^{B(n+1)}-\p_A\tilde{\xi}^{u(n-1)}-\p_A\tilde{\xi}^{r(n-1)}\\
\left[\delta \tilde{g}_{rA}\right]^{(n-1)}&=-\p_A\tilde{\xi}^{u(n-1)}-n\gamma_{AB}\tilde{\xi}^{B(n)}\\
\left[\delta \tilde{g}_{AB}\right]^{(n-2)}&=\gamma_{BC}D_A\tilde{\xi}^{C(n)}+\gamma_{AC} D_B\tilde{\xi}^{C(n)}+2\gamma_{AB}\tilde{\xi}^{r(n-1)} .
\end{array}\ee

\section{Residual Gauge Fixing}

The residual diffeomorphisms that preserve harmonic gauge~(\ref{eq:harga}) are solutions of $\Box \xi_\mu=0$.  From~(\ref{eq:vec1}-\ref{eq:vec2}) the modes 
\begin{equation}
\xi^{u(1)},~~\xi^{r(1)},~~\xi^{A(2)} 
\end{equation}
(note the raised indices) are free data that can have arbitrary $(u,z,\bz)$ dependence and preserve the boundary conditions~(\ref{eq:hfalloff}).  Subleading-in-$r$ modes are determined recursively from these using $\Box \xi_\mu=0$.  
We use the above free functions to perform residual gauge fixing that further restricts our class of large-$r$ falloffs from those of~(\ref{eq:hfalloff}).  The Lie derivative of the metric at the relevant orders of $r$ are given by~(\ref{eq:shiftpower}-\ref{eq:shiftlog})
\begin{equation}
\begin{aligned}
[\delta g_{uu}]^{(1)} &= 2 [\nabla_u \xi_u]^{(1)} = - 2 \partial_u ( \xi^{u(1)}+\xi^{r(1)} )  \\
[\delta g_{ur}]^{(1)} &= [\nabla_r \xi_u + \nabla_u \xi_r]^{(1)} = - \partial_u \xi^{u(1)}  - (\tilde{\xi}^{u(0)} + \tilde{\xi}^{r(0)})\\
[\delta g_{uA}]^{(0)} &= [\nabla_u \xi_A + \nabla_A \xi_u]^{(0)} = \gamma_{AB}\partial_u  \xi^{B(2)} - \partial_A ( \xi^{u(0)}+\xi^{r(0)} ) .
\end{aligned}
\end{equation}
We are interested in the trace reversed perturbations $h_{\mu\nu}$ appearing in $g_{\mu\nu}=\eta_{\mu\nu}+h_{\mu\nu}-\frac{1}{2}\eta_{\mu\nu}h$, where
\begin{equation}
h = h_{rr} - 2h_{ur} + \frac{1}{r^2}\gamma^{AB}h_{AB}.
\end{equation}
Taking into account the harmonic gauge condition
\begin{equation}
[\nabla^{\mu}h_{\mu u}]^{(1)} = -\partial_u h_{ur}^{(1)} = 0,~~~[\nabla^{\mu}h_{\mu r}]^{(1)} = -\partial_u h_{rr}^{(1)} = 0,
 \end{equation}
implies we can find a consistent solution where  $h_{ur}^{(1)} =h_{rr}^{(1)}= 0$.  We
then see that the $u$-dependence of $\{\xi^{u(1)},~\xi^{r(1)},~\xi^{A(2)}\}$ allows us to place the following restrictions on our trace-reversed perturbation
\be\label{eq:setzero}
h^{(1)}=\gamma^{AB}h_{AB}^{(-1)}  = 0,~~\delta g_{uu}^{(1)}-\frac{1}{2}h^{(1)}=h_{uu}^{(1)}=0,~~\delta g_{uA}^{(0)}=h_{uA}^{(0)} = 0.
\ee
The harmonic gauge condition also implies
\be
[\nabla^{\mu}h_{\mu A}]^{(0)} = -\partial_u h_{rA}^{(0)} = 0, 
\ee
so that we can consider solutions with $h_{rA}^{(0)}=0$.  We can now turn to the large-$r$ modes of the Einstein equations at orders for which the stress tensor is zero, for example
\begin{equation}
\begin{aligned}
 [\square h_{AB}]^{(0)}=-2\partial_u\tilde{h}_{AB}^{(-1)}  = 0. \\
\end{aligned}
\end{equation}
Since this mode must be $u$-independent, we can find a consistent solution where it is identically zero.   Similar considerations for $\{ [\square h_{uA}]^{(1)}, [\square h_{rA}]^{(1)},[\square h_{rr}]^{(2)},[\square h_{ur}]^{(2)}\}$ allow us to restrict to  $\{\tilde{h}_{uA}^{(0)},\tilde{h}_{rA}^{(0)},\tilde{h}_{rr}^{(1)},\tilde{h}_{ur}^{(1)}\}=0.$ Proceeding to plug in these updated falloffs into remaining modes of the Einstein tensor that vanish by~(\ref{eq:Gfalloff}), in particular $\{ [\square \tilde{h}_{rA}]^{(2)}, [\square \tilde{h}_{rr}]^{(3)},$ $[\square {h}_{rr}]^{(3)}\}$, we can set $\{\tilde{h}_{rA}^{(1)},\tilde{h}_{rr}^{(2)},{h}_{rr}^{(2)}\}=0$,
and we finally arrive at the falloffs
\begin{equation}
\begin{aligned}\label{eq:finfal1}
h_{uu} \sim \mathcal{O}(r^{-2}), \ \ \ \ \ h_{ur} \sim \mathcal{O}(r^{-2}), \ \ \ \ \ h_{rr} \sim \mathcal{O}(r^{-3}), \\
h_{uA} \sim \mathcal{O}(r^{-1}), \ \ \ \ \ h_{rA} \sim \mathcal{O}(r^{-1}), \ \ \ \ \ h_{AB} \sim \mathcal{O}(r), 
\end{aligned}
\end{equation}
and, for the log coefficients,
\begin{equation}
\begin{aligned}\label{eq:finfal2}
\tilde{h}_{uu} \sim \mathcal{O}(r^{-1}\log r), \ \ \ \ \ \tilde{h}_{ur} \sim \mathcal{O}(r^{-2} \log r), \ \ \ \ \ \tilde{h}_{rr} \sim \mathcal{O}(r^{-3} \log r), \\
\tilde{h}_{uA} \sim \mathcal{O}(r^{-1}\log r), \ \ \ \ \ \tilde{h}_{rA} \sim \mathcal{O}(r^{-2}\log r), \ \ \ \ \ \tilde{h}_{AB} \sim \mathcal{O}(\log r). 
\end{aligned}
\end{equation}

\section{Asymptotic Symmetries}
We will now consider the set of diffeomorphisms that preserve harmonic gauge as well as the falloffs~(\ref{eq:finfal1}-\ref{eq:finfal2}).  These falloffs for the trace-reversed perturbation now imply the same for $\delta g_{\mu\nu}$ because the trace is $\mathcal{O}(r^{-2}\log r)$ after our residual gauge fixing.  Since the residual gauge fixing of the previous appendix used up the $u$-dependence of the free data for a harmonic vector field to arrive at our final falloffs, we expect our solutions to be parameterized by data on the celestial sphere, functions of $(z,\bz)$.

We see from the $rr$ components of~(\ref{eq:shiftpower}-\ref{eq:shiftlog}) for $n< 3$, that $\xi^{u(n)}=0$ and $\tilde{\xi}^{u(n)}=0$ when $n<2$, with the exception of an allowed $\xi^{u(0)}$
\be
\xi^u=\xi^{u(0)}+\xi^{u(2)}r^{-2}+\tilde{\xi}^{u(2)}r^{-2}\log r+...
\ee
A similar analysis applied to metric variations, taking into account~(\ref{eq:finfal1}) and the trace condition in~(\ref{eq:setzero}) leads to
\begin{align}
\xi^u&=\frac{u}{2}D^AY_A+f+\mathcal{O}(r^{-2}\log r)\notag\\
\xi^r&=-\frac{r}{2}D^AY_A-\frac{u}{2}D^AY_A+\frac{1}{2}D^2f+Hr^{-1}+(\frac{u}{4}D^2[D^2+2]f-E)r^{-1}\log r+\mathcal{O}(r^{-2}\log r)\notag\\
\xi^B&=Y^B-D^B(\frac{u}{2}D^AY_A+f)r^{-1}+(\frac{u}{2}D^B[D^2+2]f+V^B)r^{-2}+\mathcal{O}(r^{-3}\log r)
\end{align}
where
\be\label{eq:alldiff}
\{f(z,\bz), Y^z(z), H(z,\bz), E(z,\bz), V^A(z,\bz),...\}
\ee
and we have labeled the leading terms to conform to the conventional notation used for supertranslations and superrotations, parameterized by $f(z,\bz)$ and $Y^z(z)$.  A non-holomorphic choice for $Y^z$ would modify the sphere metric at $\mathcal{O}(r^2)$.  At poles a harmonic solution can still be found as long as one relaxes our radial falloffs to hold almost everywhere on the celestial sphere.  There thus appear additional $u$-dependent delta-function supported terms which will not be relevant to our analysis and we have suppressed them here.

\bibliography{SubleadingSoftGravitonArXiv}

\providecommand{\href}[2]{#2}\begingroup\raggedright\begin{thebibliography}{10}

\bibitem{Himwich:2019dug}
E.~Himwich and A.~Strominger, ``{Celestial Current Algebra from Low's
  Subleading Soft Theorem},''
\href{http://arxiv.org/abs/1901.01622}{{\ttfamily arXiv:1901.01622 [hep-th]}}.

\bibitem{He2014}
T.~{He}, P.~{Mitra}, A.~P. {Porfyriadis}, and A.~{Strominger}, ``{New
  symmetries of massless QED},''
  \href{http://dx.doi.org/10.1007/JHEP10(2014)112}{{\em JHEP} {\bfseries 10}
  (2014) 112}, \href{http://arxiv.org/abs/1407.3789}{{\ttfamily arXiv:1407.3789
  [hep-th]}}.

\bibitem{Kapec2014}
D.~{Kapec}, V.~{Lysov}, S.~{Pasterski}, and A.~{Strominger}, ``{Semiclassical
  Virasoro symmetry of the quantum gravity $\mathcal{S}$-matrix},''
  \href{http://dx.doi.org/10.1007/JHEP08(2014)058}{{\em JHEP} {\bfseries 8}
  (2014) 58}, \href{http://arxiv.org/abs/1406.3312}{{\ttfamily arXiv:1406.3312
  [hep-th]}}.

\bibitem{Kapec2017}
D.~{Kapec}, P.~{Mitra}, A.-M. {Raclariu}, and A.~{Strominger}, ``{2D Stress
  Tensor for 4D Gravity},''
  \href{http://dx.doi.org/10.1103/PhysRevLett.119.121601}{{\em Phys. Rev.
  Lett.} {\bfseries 119} (2017) 121601},
  \href{http://arxiv.org/abs/1609.00282}{{\ttfamily arXiv:1609.00282
  [hep-th]}}.

\bibitem{Pasterski2016}
S.~{Pasterski}, A.~{Strominger}, and A.~{Zhiboedov}, ``{New gravitational
  memories},'' \href{http://dx.doi.org/10.1007/JHEP12(2016)053}{{\em JHEP}
  {\bfseries 12} (2016) 53}, \href{http://arxiv.org/abs/1502.06120}{{\ttfamily
  arXiv:1502.06120 [hep-th]}}.

\bibitem{He:2014laa}
T.~He, V.~Lysov, P.~Mitra, and A.~Strominger, ``{BMS supertranslations and
  Weinberg's soft graviton theorem},''
  \href{http://dx.doi.org/10.1007/JHEP05(2015)151}{{\em JHEP} {\bfseries 05}
  (2015) 151},
\href{http://arxiv.org/abs/1401.7026}{{\ttfamily arXiv:1401.7026 [hep-th]}}.

\bibitem{Donnay:2018neh}
L.~Donnay, A.~Puhm, and A.~Strominger, ``{Conformally Soft Photons and
  Gravitons},''
\href{http://arxiv.org/abs/1810.05219}{{\ttfamily arXiv:1810.05219 [hep-th]}}.

\bibitem{Campiglia2017}
M.~{Campiglia} and A.~{Laddha}, ``{Sub-subleading soft gravitons and large
  diffeomorphisms},'' \href{http://dx.doi.org/10.1007/JHEP01(2017)036}{{\em
  JHEP} {\bfseries 1} (2017) 36},
  \href{http://arxiv.org/abs/1608.00685}{{\ttfamily arXiv:1608.00685 [gr-qc]}}.

\bibitem{Pate2018}
M.~{Pate}, A.-M. {Raclariu}, and A.~{Strominger}, ``{Gravitational memory in
  higher dimensions},'' \href{http://dx.doi.org/10.1007/JHEP06(2018)138}{{\em
  JHEP} {\bfseries 6} (2018) 138},
  \href{http://arxiv.org/abs/1712.01204}{{\ttfamily arXiv:1712.01204
  [hep-th]}}.

\bibitem{Wald:1999wa}
R.~M. Wald and A.~Zoupas, ``{A General definition of `conserved quantities' in
  general relativity and other theories of gravity},''
  \href{http://dx.doi.org/10.1103/PhysRevD.61.084027}{{\em Phys. Rev.}
  {\bfseries D61} (2000) 084027},
\href{http://arxiv.org/abs/gr-qc/9911095}{{\ttfamily arXiv:gr-qc/9911095
  [gr-qc]}}.

\bibitem{Barnich2011}
G.~{Barnich} and C.~{Troessaert}, ``{BMS charge algebra},''
  \href{http://dx.doi.org/10.1007/JHEP12(2011)105}{{\em JHEP} {\bfseries 12}
  (2011) 105}, \href{http://arxiv.org/abs/1106.0213}{{\ttfamily arXiv:1106.0213
  [hep-th]}}.

\bibitem{Newman1962}
E.~{Newman} and R.~{Penrose}, ``{An Approach to Gravitational Radiation by a
  Method of Spin Coefficients},''
  \href{http://dx.doi.org/10.1063/1.1724257}{{\em J. Math. Phys.} {\bfseries 3}
  (1962) 566--578}.

\bibitem{Strominger2016a}
A.~{Strominger} and A.~{Zhiboedov}, ``{Gravitational memory, BMS
  supertranslations and soft theorems},''
  \href{http://dx.doi.org/10.1007/JHEP01(2016)086}{{\em JHEP} {\bfseries 01}
  (2016) 86}, \href{http://arxiv.org/abs/1411.5745}{{\ttfamily arXiv:1411.5745
  [hep-th]}}.

\bibitem{Strominger:2016wns}
A.~Strominger and A.~Zhiboedov, ``{Superrotations and Black Hole Pair
  Creation},'' \href{http://dx.doi.org/10.1088/1361-6382/aa5b5f}{{\em Class.
  Quant. Grav.} {\bfseries 34} (2017) 064002},
\href{http://arxiv.org/abs/1610.00639}{{\ttfamily arXiv:1610.00639 [hep-th]}}.

\bibitem{Strominger:2013jfa}
A.~Strominger, ``{On BMS Invariance of Gravitational Scattering},''
  \href{http://dx.doi.org/10.1007/JHEP07(2014)152}{{\em JHEP} {\bfseries 07}
  (2014) 152},
\href{http://arxiv.org/abs/1312.2229}{{\ttfamily arXiv:1312.2229 [hep-th]}}.

\bibitem{Pasterski:2017kqt}
S.~Pasterski and S.-H. Shao, ``{Conformal basis for flat space amplitudes},''
  \href{http://dx.doi.org/10.1103/PhysRevD.96.065022}{{\em Phys. Rev.}
  {\bfseries D96} (2017) 065022},
\href{http://arxiv.org/abs/1705.01027}{{\ttfamily arXiv:1705.01027 [hep-th]}}.

\bibitem{Campiglia2014}
M.~{Campiglia} and A.~{Laddha}, ``{Asymptotic symmetries and subleading soft
  graviton theorem},'' \href{http://dx.doi.org/10.1103/PhysRevD.90.124028}{{\em
  Phys. Rev.} {\bfseries D90} (2014) 124028},
  \href{http://arxiv.org/abs/1408.2228}{{\ttfamily arXiv:1408.2228 [hep-th]}}.

\bibitem{Conde:2016rom}
E.~Conde and P.~Mao, ``{BMS Supertranslations and Not So Soft Gravitons},''
  \href{http://dx.doi.org/10.1007/JHEP05(2017)060}{{\em JHEP} {\bfseries 05}
  (2017) 60},
\href{http://arxiv.org/abs/1612.08294}{{\ttfamily arXiv:1612.08294 [hep-th]}}.

\end{thebibliography}\endgroup
\bibliographystyle{utphys}

\end{document}